\documentclass[pra,nobibnotes,aps,showpieces,superscriptaddress,amsfonts,amsmath,float fix,backend=bibtex]{revtex4}
\usepackage[capitalize]{cleveref}
\usepackage{graphicx}
\usepackage{color}
\usepackage{import}
\usepackage{wrapfig}
\usepackage{mathtools}
\usepackage{bm}
\usepackage{amsmath}
\usepackage{amsfonts}
\usepackage{amssymb}

\begin{document}


\title{Addendum: Modeling the amplitude and energy decay of a weakly damped harmonic oscillator using the energy dissipation rate and a simple trick \textnormal{(2025 Eur. J. Phys. \textbf{46}(1) 015004)}}

\def\correspondingauthor{\footnote{Corresponding author}}

\author{Karlo Lelas\correspondingauthor{}}
\email{klelas@ttf.unizg.hr}
\affiliation{Faculty of Textile Technology, University of Zagreb, Prilaz baruna Filipovića 28a, 10000 Zagreb, Croatia}

\author{Robert Pezer}
\email{rpezer@simet.unizg.hr}
\affiliation{Department of Physical Metallurgy, Faculty of Metallurgy, University of Zagreb, Aleja narodnih heroja 3, 44000 Sisak, Croatia}

\date{\today}

\begin{abstract}
We show how to adapt the approach introduced for viscous damping in \cite{LelasPezer} to derive the approximate amplitude decay in the case of damping by a force of constant magnitude (sliding friction) and in the case of damping by a force proportional to the square of velocity (air resistance). We obtain two first-order differential equations from which we obtain the approximate time-dependent amplitudes corresponding to the considered damping forces. 
Our approach is suitable for first-year undergraduates, as it relies on the physical concepts and mathematical techniques they are familiar with.
\end{abstract}

\maketitle

\section{Introduction}

In a recent paper \cite{LelasPezer} viscous damping was considered and the exponential amplitude decay of a weakly damped harmonic oscillator was derived, using only basic knowledge about the undamped harmonic oscillator and the connection between the power of the damping force and the energy dissipation rate, i.e. without solving the associated equation of motion. The trick is in adding the dissipation rates corresponding to two specific pairs of initial conditions \cite{LelasPezer}. In this approach, there is no need for the time averaging used in \cite{Wang, Kontomaris2024}. 
It was pointed out in \cite{LelasPezer} that the trick of adding energy dissipation rates does not work in the case of damping by sliding friction or air resistance, i.e. it does not lead to a simple first order differential equation for the amplitude, as in the case of viscous damping. The reason for this is that the power of the damping force is not proportional to the square of the velocity for the other two types of damping, i.e. if the same approach is applied directly to the other two types of damping, one does not obtain expressions that can be simplified using trigonometric identities.

The aim of this addendum is to show how to adapt the reasoning and the trick introduced in \cite{LelasPezer} to model harmonic oscillations weakly damped by sliding friction or air resistance. This addendum is organized into six sections. In section \ref{Basic}, we present the theory and approximations needed in our approach. In section \ref{sliding friction}, we derive the amplitude decay in case of sliding friction. In section \ref{air resistance}, we derive the amplitude decay in case of air resistance. In section \ref{comparison1}, we compare our approximate solution in case of sliding friction with the approximate solution obtained in \cite{Wang} and with the exact solution given in \cite{Grk2}. In section \ref{comparison2}, we compare our approximate solution in case of air resistance with the results obtained in \cite{Wang}. In section \ref{conclusion}, we comment on some issues with the choices of initial conditions in our approach, and summarize the presented results.

\section{Basic theory and approximations we use}
\label{Basic}

As in \cite{LelasPezer}, we consider a block of mass $m$ that oscillates under the influence of the restoring force of an ideal spring of stiffness $k$ and the damping force $F_d(t)$. The equation of motion of this system is 
\begin{equation}
ma(t)=F_d(t)-kx(t)\,,
\label{HOeq}
\end{equation}
where $x(t)$ is the block’s displacement from the equilibrium position, and $a(t)=d^2x(t)/dt^2$ is its acceleration. 
%
%
Damping by sliding friction can be modeled with \cite{Lapidus, AviAJP, Grk2}
\begin{equation}
F_d(t)=-\text{sgn}[v(t)]\,\mu mg\,,
\label{slidingF}
\end{equation} 
where $v(t)=dx(t)/dt$ is the velocity of the block, $\mu>0$ is the coefficient of sliding friction \cite{Young2020university,Resnick10}, and we use the sign function
\begin{equation} \label{signum}
    \text{sgn}\left[v(t)\right] = \begin{cases}
\begin{tabular}{@{}cl@{}}
   $1$\, & if\, $v(t)$ $ > $ $0$ \\
    $0$\, & if\, $v(t)$ $ = $ $0$ \\
    $-1$\, & if\, $v(t)$ $ < $ $0$\, 
\end{tabular}
    \end{cases}
    \end{equation}
to take into account that damping force always has a direction opposite to the direction of the velocity. Damping by air resistance can be modeled with
\begin{equation}
F_d(t)=-\text{sgn}[v(t)]\,Cv^2(t)\,,
\label{airF}
\end{equation}
where $C>0$ is a constant \cite{Young2020university, Resnick10}. In the context of air resistance, we can imagine that we have a sphere of mass $m$ attached to a spring, rather than the block, since usually air resistance of spherical objects is considered in the physics textbooks \cite{Young2020university, Resnick10}, and in that case $C=c\rho A/2$, where $c$ is experimentally determined drag coefficient, $\rho$ is the air density, and $A$ is the cross sectional area of the sphere \cite{Resnick10}. 
The energy (potential plus kinetic) of the block-spring system is given by  
\begin{equation}
E(t)=\frac{kx^2(t)}{2}+\frac{mv^2(t)}{2}\,.
\label{Energy}
\end{equation}
%
The energy is not conserved due to the power of the damping force $P_d(t)=F_d(t)v(t)$, and the energy dissipation rate is given by \cite{Young2020university} 
\begin{equation}
\frac{dE(t)}{dt}=F_d(t)v(t)\,.
\label{Power}
\end{equation}
Equation \eqref{Power} is valid for both types of damping, i.e for damping forces \eqref{slidingF} and \eqref{airF}. Since the direction of the damping force is opposite to the direction of the velocity at all times, $F_d(t)v(t)\leq0$ is valid $\forall t$. Thus, equation \eqref{Power} tells us that the energy of the damped system decreases monotonically with time.


Two pairs of initial conditions, i.e. $(x_1(0)=x_0, v_1(0)=0)$ and $(x_2(0)=0, v_2(0)=\omega_0x_0)$, were considered in \cite{LelasPezer}. The first pair has purely potential initial energy, and the second pair has purely kinetic initial energy, and, due to the chosen value of $v_2(0)$, both pairs have the same initial energy $E_0=m\omega_0^2x_0^2/2$. The solutions of equation \eqref{HOeq} in the undamped case, corresponding to these initial conditions, are $x_1(t)=x_0\cos(\omega_0t)$ and $x_2(t)=x_0\sin(\omega_0t)$.  
Thus, similarly as in the case of weak viscous damping presented in \cite{LelasPezer}, we can take that the solutions of equation \eqref{HOeq}, with weak damping force \eqref{slidingF} or \eqref{airF}, are approximately of the form 
\begin{equation}
x_1(t)=x_0f(t)\cos(\omega_0t)
\label{Xansatz1}
\end{equation}
for the first pair of initial conditions, and
\begin{equation}
x_2(t)=x_0f(t)\sin(\omega_0t)\,
\label{Xansatz2}
\end{equation}
for the second pair of initial conditions, where $f(t)$ is the unknown function that describes the decrease in amplitude over time. We can take the velocities, corresponding to \eqref{Xansatz1} and \eqref{Xansatz2}, to be approximately of the form 
\begin{equation}
v_1(t)=-\omega_0x_0f(t)\sin(\omega_0t)\,
\label{Vansatz11}
\end{equation}
and
\begin{equation}
v_2(t)=\omega_0x_0f(t)\cos(\omega_0t)\,,
\label{Vansatz22}
\end{equation}
due to weak damping condition $|\dot{f}(t)|\ll\omega_0$ \cite{LelasPezer} (where dot denotes the time derivative). It is easy to see that $f(0)=1$ must hold for the functions \eqref{Xansatz1} and \eqref{Vansatz22} to satisfy the initial conditions $x_1(0)=x_0$ and $v_2(0)=\omega_0x_0$. Since $f(0)=1$ holds and the function monotonically decreases, we can easily conclude that $f(t)\geq0$ must also hold, for all $t\geq0$. Using approximate displacements \eqref{Xansatz1} and \eqref{Xansatz2}, and approximate velocities \eqref{Vansatz11} and \eqref{Vansatz22} in \eqref{Energy}, we get the corresponding approximate energies
\begin{equation}
E_1(t)=E_2(t)=\frac{m\omega_0^2x_0^2f^2(t)}{2}\,.
\label{Energy12}
\end{equation}
The first derivatives of the energies \eqref{Energy12} are
\begin{equation}
\frac{dE_1(t)}{dt}=\frac{dE_2(t)}{dt}=m\omega_0^2x_0^2f(t)\frac{df(t)}{dt}\,.
\label{derivEnergy12}
\end{equation}

\section{Amplitude decay in case of weak sliding friction}
\label{sliding friction}


In case of weak damping with $F_d(t)=-\text{sgn}[v(t)]\mu mg$ we have 
\begin{equation}
P_{d1}(t)=-\text{sgn}[v_1(t)]\mu mg\,v_1(t)\,
\label{slidingP1}
\end{equation}
as the power of the damping force corresponding to the first pair of initial conditions, and
\begin{equation}
P_{d2}(t)=-\text{sgn}[v_2(t)]\mu mg\,v_2(t)\,
\label{slidingP2}
\end{equation}
as the power of the damping force corresponding to the second pair of initial conditions. It is easy to show that $\text{sgn}[v(t)] v(t)=|v(t)|$ is valid. We take \eqref{Vansatz11} and \eqref{Vansatz22} for velocities $v_1(t)$ and $v_2(t)$. Thus, we can approximate \eqref{slidingP1} and \eqref{slidingP2} with 
\begin{equation}
P_{d1}(t)=-\mu mg\omega_0x_0f(t)|\sin(\omega_0t)|\,
\label{AslidingP1}
\end{equation}
and
\begin{equation}
P_{d2}(t)=-\mu mg\omega_0x_0f(t)|\cos(\omega_0t)|\,.
\label{AslidingP2}
\end{equation}
Using \eqref{derivEnergy12} to approximate the left hand side of $dE(t)/dt=P_{d}(t)$, i.e. of the energy dissipation rate \eqref{Power}, and relations \eqref{AslidingP1} and \eqref{AslidingP2} to approximate the right hand side of \eqref{Power}, we get
\begin{equation}
\frac{df(t)}{dt}=-\frac{\mu g}{\omega_0x_0}\,|\sin(\omega_0t)|\,
\label{slidingF1}
\end{equation}
for the first pair of initial conditions, and
\begin{equation}
\frac{df(t)}{dt}=-\frac{\mu g}{\omega_0x_0}\,|\cos(\omega_0t)|\,
\label{slidingF2}
\end{equation}
for the second pair of initial conditions. We can now take the square of relations \eqref{slidingF1} and \eqref{slidingF2}, we get
\begin{equation}
\left(\frac{df(t)}{dt}\right)^2=\left(\frac{\mu g}{\omega_0x_0}\right)^2\,|\sin(\omega_0t)|^2\,
\label{slidingF11}
\end{equation}
and
\begin{equation}
\left(\frac{df(t)}{dt}\right)^2=\left(\frac{\mu g}{\omega_0x_0}\right)^2\,|\cos(\omega_0t)|^2\,
\label{slidingF22}
\end{equation}
By adding \eqref{slidingF11} and \eqref{slidingF22} we get
\begin{equation}
\left(\frac{df(t)}{dt}\right)^2=\frac{1}{2}\left(\frac{\mu g}{\omega_0x_0}\right)^2\,,
\label{slidingF12}
\end{equation}
since identity $|\sin(\omega_0t)|^2+|\cos(\omega_0t)|^2=1$ is valid. By taking the square root of \eqref{slidingF12} we obtain
\begin{equation}
\frac{df(t)}{dt}=-\frac{\mu g}{\sqrt{2}\,\omega_0 x_0}\,.
\label{slidingF12final}
\end{equation}
We take the negative value of the square root in \eqref{slidingF12final}, since $f(t)$ has to be a monotonically decreasing function. Even without knowing integral calculus, it is easy to conclude that the solution of \eqref{slidingF12final} is $f(t)=-\frac{\mu g}{\sqrt{2}\,\omega_0 x_0}t+const$. The constant term is determined from the condition $f(0)=1$. Thus, the final expression for the amplitude decay is
\begin{equation}
f(t)=1-\frac{\mu g}{\sqrt{2}\,\omega_0 x_0}t\,.
\label{slidingFfinal}
\end{equation}
Thus, our approximate solutions and energies, in case of sliding friction, are obtained by inserting \eqref{slidingFfinal} in \eqref{Xansatz1}, \eqref{Xansatz2} and \eqref{Energy12}. The relation \eqref{slidingFfinal} tells us that the system stops in the equilibrium position at instant $\tau=\sqrt{2}\omega_0x_0/(\mu g)$, because at that moment both displacements \eqref{Xansatz1} and \eqref{Xansatz2}, and velocities \eqref{Vansatz11} and \eqref{Vansatz22}, are equal to zero, and for $t>\tau$ our approximate solutions are no longer physical, i.e. the amplitude becomes negative and increase in magnitude for $t>\tau$. In addition, we can now easily find what condition must be satisfied in order to quantify that we are in the regime of weak damping with sliding friction. As we stated before, for weak damping condition $|\dot{f}(t)|\ll\omega_0$ holds. In case of \eqref{slidingFfinal} we have $|\dot{f}(t)|=\frac{\mu g}{\sqrt{2}\,\omega_0 x_0}$ for all $t\geq0$. Thus, if we use $\omega_0=\sqrt{k/m}$ we get 
\begin{equation}
\mu\ll\frac{\sqrt{2}\,kx_0}{mg}\,.
\label{uvjet1}
\end{equation}
as the condition of weak damping in the case of sliding friction.

\section{Amplitude decay in case of weak air resistance}
\label{air resistance}

In case of weak damping with $F_d(t)=-\text{sgn}[v(t)]\,Cv^2(t)$, the expressions for the power of the damping force are
\begin{equation}
P_{d1}(t)=-\text{sgn}[v_1(t)]Cv_1^3(t)\,
\label{airP1}
\end{equation}
and
\begin{equation}
P_{d2}(t)=-\text{sgn}[v_2(t)]Cv_2^3(t)\,.
\label{airP2}
\end{equation}
It is easy to show that $\text{sgn}[v(t)] v^3(t)=|v(t)|^3$ is valid. We take \eqref{Vansatz11} and \eqref{Vansatz22} for velocities $v_1(t)$ and $v_2(t)$. Thus, we can approximate \eqref{airP1} and \eqref{airP2} with 
\begin{equation}
P_{d1}(t)=-C\left(\omega_0x_0f(t)\right)^3|\sin(\omega_0t)|^3\,
\label{airP11}
\end{equation}
and
\begin{equation}
P_{d2}(t)=-C\left(\omega_0x_0f(t)\right)^3|\cos(\omega_0t)|^3\,
\label{airP22}
\end{equation}
Using \eqref{derivEnergy12} to approximate the left hand side of the energy dissipation rate \eqref{Power}, and relations \eqref{airP11} and \eqref{airP22} to approximate the right hand side of \eqref{Power}, i.e. by approximating the energy dissipation rate $dE(t)/dt=P_{d}(t)$ for both pairs of initial conditions, we get 
\begin{equation}
-\frac{df(t)}{dt}=\frac{C\omega_0x_0}{m}f^2(t)|\sin(\omega_0t)|^3\,
\label{airF1}
\end{equation}
and
\begin{equation}
-\frac{df(t)}{dt}=\frac{C\omega_0x_0}{m}f^2(t)|\cos(\omega_0t)|^3\,.
\label{airF2}
\end{equation}
The left and right sides of \eqref{airF1} and \eqref{airF2} are positive for any $t$. We raise the relations \eqref{airF1} and \eqref{airF2} to the power of $2/3$ and get 
\begin{equation}
\left(-\frac{df(t)}{dt}\right)^{\frac{2}{3}}=\left(\frac{C\omega_0x_0}{m}\right)^{\frac{2}{3}}f^{\frac{4}{3}}(t)|\sin(\omega_0t)|^2\,
\label{airF11}
\end{equation}
and
\begin{equation}
\left(-\frac{df(t)}{dt}\right)^{\frac{2}{3}}=\left(\frac{C\omega_0x_0}{m}\right)^{\frac{2}{3}}f^{\frac{4}{3}}(t)|\cos(\omega_0t)|^2\,
\label{airF22}
\end{equation}
We add \eqref{airF11} and \eqref{airF22} and get 
\begin{equation}
\left(-\frac{df(t)}{dt}\right)^{\frac{2}{3}}=\frac{1}{2}\left(\frac{C\omega_0x_0}{m}\right)^{\frac{2}{3}}f^{\frac{4}{3}}(t)\,.
\label{airF12}
\end{equation}
We raise the relation \eqref{airF12} to the power of $3/2$ and get
\begin{equation}
\frac{df(t)}{dt}=-\left(\frac{C\omega_0x_0}{2^{\frac{3}{2}}m}\right)f^2(t)\,.
\label{airF122}
\end{equation}
We perform integration
\begin{equation}
\int_{f(0)}^{f(t)}f^{-2}(t')df(t')=-\frac{C\omega_0x_0}{2^{\frac{3}{2}}m}\int_0^tdt'
\label{integral}
\end{equation}
and finally obtain  
\begin{equation}
f(t)=\left(1+\frac{C\omega_0x_0}{2^{\frac{3}{2}}m}t\right)^{-1}\,.
\label{airF123}
\end{equation}
We note here that integrals of the form \eqref{integral} are known to first-year undergraduates \cite{Young2020university, Resnick10}. Thus, our approximate solutions and energies, in case of air resistance, are obtained by inserting \eqref{airF123} into \eqref{Xansatz1}, \eqref{Xansatz2} and \eqref{Energy12}. 
As we stated before, condition $|\dot{f}(t)|\ll\omega_0$ holds in the case of weak damping. This condition holds for all $t\geq0$, thus, if we consider $t=0$, for which $\dot{f}(0)=-C\omega_0x_0/(2^{\frac{3}{2}}m)$, we get 
\begin{equation}
C\ll\frac{2^{\frac{3}{2}}m}{x_0}
\label{uvjet2}
\end{equation}
as the condition of weak damping in the case of air resistance.

\section{Sliding friction: Comparison of our results with the results of a known method and exact results}
\label{comparison1}

In this section, we deal with the dynamics that started with initial conditions $(x(0)=x_0, v(0)=0)$. In that case, our approximate solution of equation \eqref{HOeq}, with damping force \eqref{slidingF}, is
\begin{equation}
x(t)=x_0\left(1-\frac{\mu g}{\sqrt{2}\,\omega_0 x_0}t\right)\cos(\omega_0t)\,.
\label{XC}
\end{equation}
In \cite{Wang}, the amplitude decay in the case of weak damping with sliding friction is derived using the approximation that the amplitude remains constant over time intervals of one period and using the energy dissipation rate averaged over these time intervals. The method introduced in \cite{Wang} leads to approximate solution
\begin{equation}
\tilde{x}(t)=x_0\left(1-\frac{2\mu g}{\pi\,\omega_0 x_0}t\right)\cos(\omega_0t)\,
\label{XW}
\end{equation}
if applied to the system we consider here. On the other hand, equation of motion \eqref{HOeq}, with damping force \eqref{slidingF}, can be solved exactly \cite{Grk2}, but the derivation of the solution is quite demanding for first-year undergraduates. The exact solution is, see e.g. \cite{Grk2, Kamela}, 
\begin{equation}
x_{ex}(t)=\left(x_0-(2n-1)\frac{\mu mg}{k}\right)\cos(\omega_0t)-(-1)^n\frac{\mu mg}{k}\,
\label{Xex}
\end{equation}
where $n\geq1$ is an integer that represents the number of half periods, i.e. for $0\leq t\leq T_0/2$ we take $n=1$, for $T_0/2\leq t\leq T_0$ we take $n=2$, etc., where $T_0=2\pi/\omega_0$. For simplicity, here we assume that the static and dynamic coefficients of friction are the same, i.e. both equal to $\mu$. Our goal here is only to quantify the validity of our approach, so we will not engage in a detail discussion of the case when these coefficients are different and in detail discussions about where and when the exact solution \eqref{Xex} comes to a halt, it has already been thoroughly covered elsewhere, see e.g. \cite{Lapidus, AviAJP, Coulomb, Grk2}.

\begin{figure}[h!t!]
\begin{center}
\includegraphics[width=0.48\textwidth]{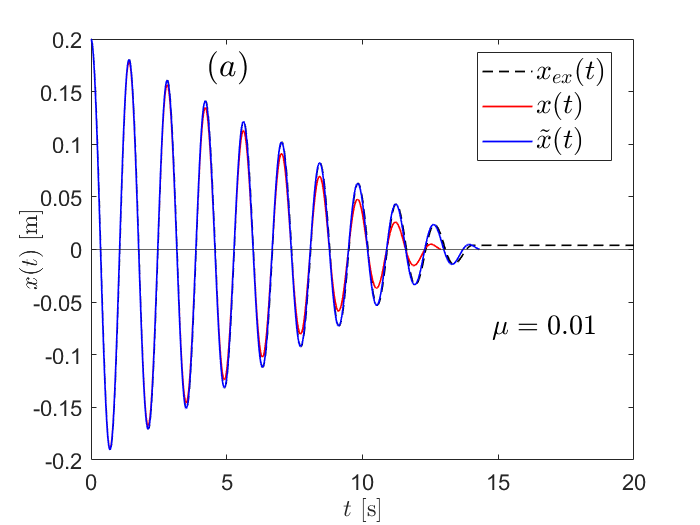}
\includegraphics[width=0.48\textwidth]{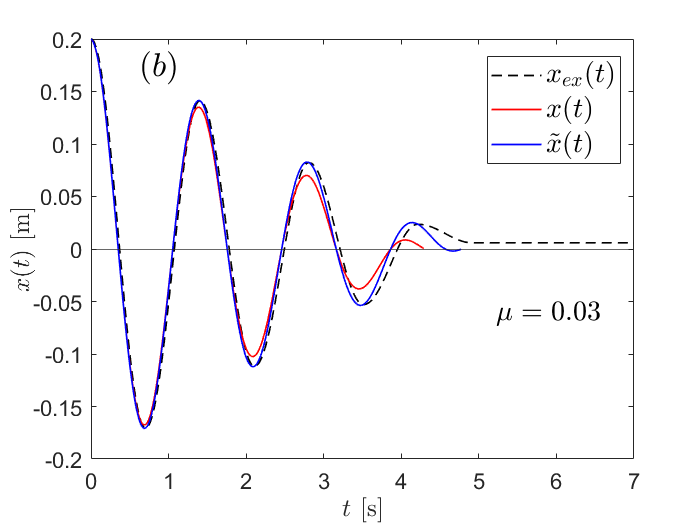}
\end{center}
\caption{Solutions \eqref{XC} (solid red curve), \eqref{XW} (solid blue curve) and \eqref{Xex} (dashed black curve), for (a) $\mu=0.01$ and (b) $\mu=0.03$. See text for details.} 
\label{sliding}
\end{figure}

In Fig.\,\ref{sliding} we show the approximate solutions \eqref{XC} (solid red curve) and \eqref{XW} (solid blue curve), and the exact solution \eqref{Xex} (dashed black curve). In both Fig.\,\ref{sliding}(a) and (b), we consider block-spring system with mass $m=1.5\,kg$, spring stiffness $k=30\,N/m$, initial displacement $x_0=0.2\,m$ and we use $g=9.81\,m/s^2$. 
For the chosen values, condition \eqref{uvjet1} becomes
\begin{equation}
\mu\ll 0.58\,.
\label{uvjet11}
\end{equation}
In Fig.\,\ref{sliding}(a) we show the dynamics with $\mu=0.01$, and in Fig.\,\ref{sliding}(b) with $\mu=0.03$.
%
%
%

As we noted earlier, solution \eqref{XC} halts in equilibrium position at $\tau=\sqrt{2}\omega_0x_0/(\mu g)$, which in this case amounts to $\tau=12.9\,s$ for $\mu=0.01$, and to $\tau=4.3\,s$ for $\mu=0.03$. In that respect, solution \eqref{XW} behaves qualitatively the same, i.e. it comes to a stop in equilibrium position at $\tilde{\tau}=\pi\omega_0x_0/(2\mu g)$, which amounts to $\tilde{\tau}=14.33\,s$ for $\mu=0.01$, and to $\tilde{\tau}=4.78\,s$ for $\mu=0.03$. We read directly from the graph that the exact solution stops at time $\tau_{ex}=14.05\,s$ at displacement $+0.02x_0$ for $\mu=0.01$, and at time $\tau_{ex}=4.93\,s$ at displacement $+0.03x_0$ for $\mu=0.03$. 

We see in Fig.\,\ref{sliding} that the solution \eqref{XW} is a better approximation of the exact solution than our solution \eqref{XC}, especially as the dynamics approaches to a halt. Nevertheless, our solution also describes well the dynamics of the systems if condition \eqref{uvjet11} is valid, i.e. our solution shows a linear decay of the amplitude and provides a solid estimate of the duration of the motion. The benefit of our approach is in a less demanding derivation compared to the approach presented in \cite{Wang}, i.e. even students who have not studied integral calculus can understand it.

\section{Air resistance: Comparison of our results with the results of a known method and numerical results}
\label{comparison2}


As in previous section, here we deal with the dynamics that started with initial conditions $(x(0)=x_0, v(0)=0)$. In that case, our approximate solution of equation \eqref{HOeq}, with damping force \eqref{airF}, is
\begin{equation}
x(t)=x_0\left(1+\frac{C\omega_0x_0}{2^{\frac{3}{2}}m}t\right)^{-1}\cos(\omega_0t)\,.
\label{XA}
\end{equation}
Equation of motion \eqref{HOeq} with damping force \eqref{airF} is a nonlinear differential equation that cannot be solved analytically. In \cite{Wang}, amplitude decay in case of weak damping with damping force quadratic in velocity is derived using the approximation that the amplitude remains constant over time intervals of one period in length and using the energy dissipation rate averaged (integrated) over these time intervals. In \cite{Wang}, it was also shown that the approximate solution thus obtained agrees very well with the experimental data. Therefore, to check the quality of our approach in case of damping quadratic in velocity, we will compare solution \eqref{XA} with the solution given by the approach used in \cite{Wang} when applied to the block-spring system we study here, i.e. compare \eqref{XA} with
\begin{equation}
\tilde{x}(t)=x_0\left(1+\frac{4C\omega_0x_0}{3\pi m}t\right)^{-1}\cos(\omega_0t)\,.
\label{XAW}
\end{equation}

\begin{figure}[h!t!]
\begin{center}
\includegraphics[width=0.48\textwidth]{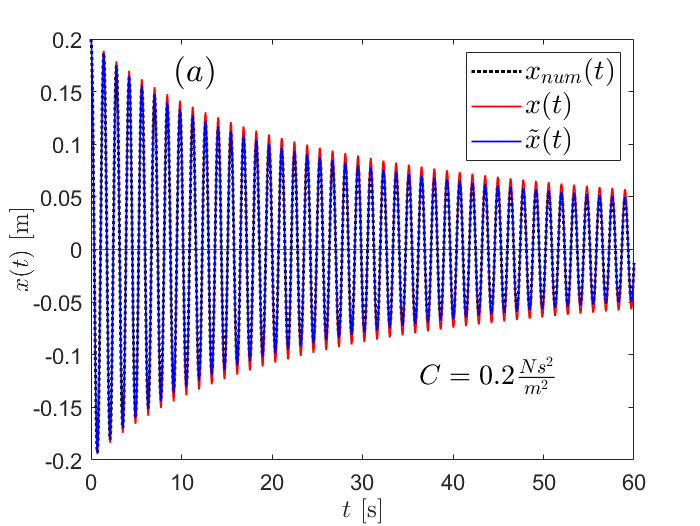}
\includegraphics[width=0.48\textwidth]{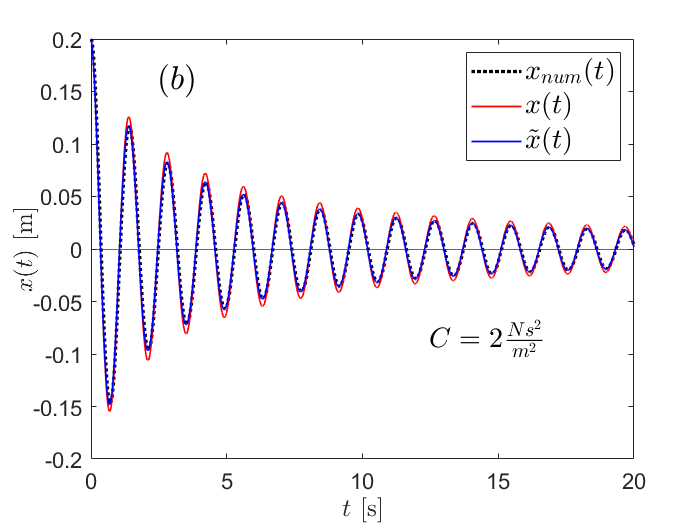}
\end{center}
\caption{Solutions \eqref{XA} (solid red curve) and \eqref{XAW} (solid blue curve), for (a) $C=0.2\,Ns^2/m^2$ and (b) $C=2\,Ns^2/m^2$. In both figures, the dotted black curves show the corresponding numerical solutions. See text for details.} 
\label{air}
\end{figure}

In Fig.\,\ref{air} we show the approximate solutions \eqref{XA} (solid red curves) and \eqref{XAW} (solid blue curves) along with the corresponding numerical solutions of the equation of motion (dotted black curves). The \emph{ode45} MATLAB function has been utilized to obtain numerical solutions of equation \eqref{HOeq} with damping force \eqref{airF}. In both Fig.\,\ref{air}(a) and (b), we use the same values of $m$, $k$, $x_0$ and $g$ as in previous section. For the chosen values, condition \eqref{uvjet2} becomes
\begin{equation}
C\ll 21.21\frac{Ns^2}{m^2}\,.
\label{uvjet22}
\end{equation}
In Fig.\,\ref{air}(a) we show the dynamics with $C=0.2\,Ns^2/m^2$, and in Fig.\,\ref{air}(b) with $C=2\,Ns^2/m^2$.

We see in Fig.\,\ref{air} that our solution \eqref{XA} agrees well with the solution \eqref{XAW} and with the numerical solutions, i.e. it gives only a slightly slower amplitude decay compared to them. Therefore, along with the approach presented in \cite{Wang}, our approach can also be used to introduce undergraduate students with the damping quadratic in velocity. In terms of mathematical complexity, a small advantage of our approach, compared to the approach in \cite{Wang}, is that students do not need to calculate two integrals, but only one, the one they are typically familiar with in the first year of undergraduate studies. More precisely, for damping quadratic in velocity the approach used in \cite{Wang} requires solving integrals $\int_0^{T_0}|\sin(\omega_0t)|^3dt$ and $\int_{f(0)}^{f(t)}f^{-2}(t')df(t')$, while in our approach we do not need to calculate the first of those two integrals.

\section{Discussion and conclusion}
\label{conclusion}

Here, we would like to comment more generally on the pairs of initial conditions which can be used in our approach. Let us take that $x_1(t)=A_0\cos(\omega_0t+\phi_0)$ and $x_2(t)=B_0\cos(\omega_0t+\beta_0)$ are the undamped solutions of equation \eqref{HOeq} corresponding to the initial conditions $\left(x_1(0)=A_0\cos(\phi_0), v_1(0)=-\omega_0A_0\sin(\phi_0)\right)$ and $\left(x_2(0)=B_0\cos(\beta_0), v_2(0)=-\omega_0B_0\sin(\beta_0)\right)$. In the case of weak damping we can take $x_1(t)=A_0f(t)\cos(\omega_0t+\phi_0)$ and $v_1(t)=-\omega_0A_0f(t)\sin(\omega_0t+\phi_0)$ as approximate forms of displacement and velocity corresponding to the first pair of initial conditions, and $x_2(t)=B_0f(t)\cos(\omega_0t+\beta_0)$ and $v_2(t)=-\omega_0B_0f(t)\sin(\omega_0t+\beta_0)$ for the second pair. If we insert these approximate forms of displacements and velocities, together with $F_d(t)\propto v^n(t)$, into the energy dissipation rate $dE(t)/dt=F_d(t)v(t)$ we get
\begin{equation}
m\omega_0^2A_0^2f(t)\frac{df(t)}{dt}\propto \omega_0^{n+1}A_0^{n+1}f^{n+1}(t)\sin^{n+1}(\omega_0t+\phi_0)\,.
\label{josmalo}
\end{equation}
\begin{equation}
m\omega_0^2B_0^2f(t)\frac{df(t)}{dt}\propto \omega_0^{n+1}B_0^{n+1}f^{n+1}(t)\sin^{n+1}(\omega_0t+\beta_0)\,.
\label{josmalo2}
\end{equation}
We see that in the case of viscous damping with $n=1$ (studied in \cite{LelasPezer}), amplitudes $A_0$ and $B_0$ can be eliminated (crossed) from equations \eqref{josmalo} and \eqref{josmalo2}. Thus, we would like to note here that in case $n=1$ the trick of adding two energy dissipation rates, introduced in \cite{LelasPezer}, can work if we use any two pairs of initial conditions that are phase shifted by $\pi/2$, i.e. if we get equation \eqref{josmalo} for the first pair and equation \eqref{josmalo2} for the other pair, the only necessary condition for the derivation introduced in \cite{LelasPezer} to work is that $\beta_0=\phi_0\pm\pi/2$, but $A_0$ and $B_0$ can be different (i.e. initial energies can be different). However, when students encounter this topic for the first time, in our opinion, it is better to explain the entire procedure to them using two pairs of initial conditions that have the same initial energy, i.e. with $A_0=B_0$, and with $\phi_0=0$ and $\beta_0=-\pi/2$, for simplicity, as is done in \cite{LelasPezer}. For other two types of damping ($n=0$ and $n=2$) the amplitudes $A_0$ and $B_0$ cannot be eliminated from equations \eqref{josmalo} and \eqref{josmalo2}, and our derivations work only if we consider pairs of initial conditions with the same initial energy, i.e. with $A_0=B_0$, and with $\beta_0=\phi_0\pm\pi/2$. 

Furthermore, the weak damping conditions \eqref{uvjet1} and \eqref{uvjet2} both depend on the initial displacement $x_0$, i.e. on initial energy, therefore the initial conditions play an important role in the overall dynamics in the case of sliding friction and air resistance, while the initial conditions do not appear in the weak damping condition $b/(2m)\ll\omega_0$ in the case of viscous damping \cite{LelasPezer}. However, we note here that the initial conditions can also play an important role in the case of viscous damping if all values of the damping constant $b$ have to be considered, not only the weak regime, as is the case in the context of optimal damping problems \cite{Lelas2023, Lelas2024}.


In conclusion, we showed how to obtain a fairly good approximation of the amplitude decay, i.e., of the solutions that describe the dynamics of harmonic oscillator damped with sliding friction or with air resistance. In case of sliding friction, we arrive at the final expression for the amplitude decay using simpler mathematics than is required for the exact solution, and there is no need for integral calculus as in \cite{Wang}. Thus, for this type of damping our approach is suitable for both first year undergraduates and for high school students. In case of air resistance, we arrive at the final expression for the amplitude decay without the need for averaging over time, i.e. in our approach, we use one integration less compared to the approach in \cite{Wang}. In this sense, our approach is somewhat easier from the mathematical side than the approach used in \cite{Wang}. The analysis of the influence of these forces on the dynamics of the harmonic oscillator is omitted in physics textbooks for the first year of undergraduate studies, e.g., see \cite{Cutnell8, Resnick10, Young2020university}. Furthermore, even in physics textbooks that give a more thorough analysis of vibrations at a more advanced undergraduate level, the influence of sliding friction and air resistance on the dynamics of oscillating systems is not considered, e.g., see \cite{Waves}. Due to its mathematical simplicity, the approach presented in this addendum can be useful in providing first-year undergraduates with insight into the dynamics of a harmonic oscillator weakly damped by these forces.

\section{Acknowledgments}

This work was supported by the QuantiXLie Center of Excellence, a project co-financed by the Croatian Government and European Union through the European Regional Development Fund, the Competitiveness and Cohesion Operational Programme (Grant No. KK.01.1.1.01.0004).

\bibliographystyle{unsrt}
\bibliography{DHO.bib}

\end{document}